\documentclass[twocolumn,secnumarabic,amssymb, amsmath,nobibnotes]{revtex4}

\usepackage{amsmath}
\usepackage{graphicx}

\begin{document}

\title{Faraday filtering of Raman light in an isotopically pure alkali metal vapor}


\author{Richard P. Abel$^*$, Ulrich Raitzsch, Paul Siddons, Ifan G. Hughes and Charles S. Adams}

\address{
Department of Physics, Durham University, \\  Rochester Building, South Road, Durham DH1 3LE, United Kingdom
\\
$^*$Corresponding author: r.p.abel@dur.ac.uk
}

\begin{abstract}We present an application of the Faraday effect to produce a narrow band atomic filter in an alkali metal vapor. In our experiment two Raman beams separated in frequency by the ground state hyperfine splitting in $^{87}$Rb are produced using an EOM and then filtered using the Faraday effect in an isotopically pure $^{85}$Rb thermal vapor. An experimental transmission spectra for the filter is presented along with a theoretical calculation. The performance of the filter is then demonstrated and characterized using a Fabry-Perot etalon. For a temperature of 70$^{\circ}$C and a longitudinal magnetic field of 80 G a suppression to -18 dB is achieved, limited by the quality of the polarizers.

\end{abstract}

\maketitle


\noindent In many atomic physics experiments one requires phase coherent laser light at frequencies separated by the ground state hyperfine splitting; examples include stimulated Raman transitions \cite{Chu91}; coherent population trapping \cite{Arimondo96}; $\Lambda$-system and $\mathcal{N}$-system electromagnetically induced transparency (EIT) \cite{Bason08,Bason09} and mesoscopic Rydberg gates \cite{Muller08}. Alternatively, Raman light is also produced in experiments involving Raman scattering processes \cite{Manz07}. In many cases the two components of the Raman light are of unequal intensity and are not spatially separated.  
We therefore require a filter that will separate the two frequencies into separate beams, producing two sources of light suitable for subsequent applications. 

A narrow band atomic filter can be realized using the Faraday effect where a longitudinal magnetic field induces a circular birefringence in the medium \cite{Faraday}. Atomic filters exploiting birefringence were first introduced and demonstrated by \"{O}hman \cite{Ohman56}. This principle was developed into the Faraday anomalous-dispersion optical filter (FADOF), which has been demonstrated in Cs \cite{Menders91}, Rb \cite{Shay91, Junxiong95, Popescu04} and Na \cite{Chen93}. Similarly, the induced-dichroism excited atomic line (IDEAL) filter, which operates without a magnetic field, has been demonstrated in K \cite{Gayen95}. More recently, atomic filters have been produced using absorption in a thermal vapor cell \cite{Manz07} and velocity selective optical pumping in an atomic vapor \cite{Mitchell09}. Narrow band atomic filters have applications in free space laser communications \cite{Junxiong95}, atmospheric measurements using LIDAR \cite{Gardiner04,Chen93}, ocean temperature profiling using LIDAR \cite{Popescu04} and the generation of narrow band quantum-noise-limited light \cite{Menders91}.  In our experiment we use an isotopically pure cell such that we can exploit the Faraday effect in $^{85}$Rb to filter Raman light resonant with $^{87}$Rb.


The Faraday effect is observed when a magnetic field is applied parallel to the direction of light propagation, causing initially linearly polarized light to be rotated by an angle, $\theta$, given by:

\begin{equation}
\theta= VBL,
\label{eq:verdet}
\end{equation}

\noindent where $V$ is the Verdet constant, $B$ the magnitude of the applied magnetic field and $L$ the length of the medium. The Verdet constant is dependent on the properties of the medium, the wavelength of the light and the temperature. Typical commercial Faraday isolators employ a Terbium Gallium garnet crystal with a Verdet constant of $134$ Rad T$^{-1}$ m$^{-1}$ at 632 nm \cite{Villaverde78}.

\begin{figure}[rb]
\centerline{\includegraphics[]{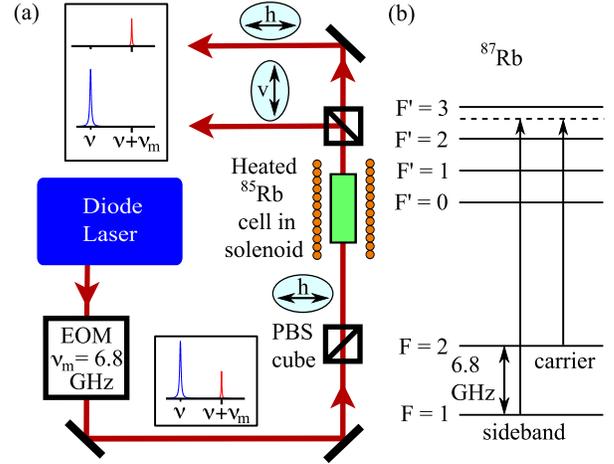}}
\caption{(a) Schematic experimental setup. An EOM is used to create sidebands at $\nu\pm\nu_m$ with $\nu_m= 6.8$ GHz producing light corresponding to the Raman transitions in $^{87}$Rb as shown in (b). The components at $\nu$ and $\nu+\nu_m$ are separated using the Faraday effect in a heated $^{85}$Rb vapor cell within a solenoid. The input light is initially horizontally polarized using a polarizing beam splitting (PBS) cube. Following the solenoid the different frequency components are separated using a second PBS cube.  The insets show  schematics of Fabry-Perot etalon signals. }
\label{fig:setup}
\end{figure}

The polarization rotation is produced by an induced circular birefringence in the medium \cite{Siddons09} which produces a relative phase shift, $\Delta\phi$, between circular field components given by:

\begin{equation}
\Delta\phi = \frac{\omega L}{c}(n^+-n^-),
\label{equation:pol_rot}
\end{equation}

\noindent where $n^{+}$ and $n^{-}$ are the refractive index of the circular components driving $\sigma^+$ and $\sigma^-$ transitions, respectively. The relative phase shift has a spectral dependance which can be expressed as: 

\begin{equation}
\frac{{\rm d}(\Delta\phi)}{{\rm d}\omega} = \frac{L}{c}(n_g^{+}-n_g^{-}).
\end{equation}

\noindent The relative phase shift causes a rotation of the polarization of initially linearly polarized light by an amount $\theta=\Delta\phi/2$ (see Methods section in \cite{Siddons09}). By exploiting the Faraday effect in an isotopically pure Rb thermal vapor it is possible to design a system with a frequency dependent polarization rotation. Light that is resonant with the lower ground state hyperfine to excited state transition in one isotope, but is in the absorption wings of the filter isotope, will be transmitted unchanged. Whereas light resonant with the upper hyperfine to excited state transition experiences a rotation in polarization equal to $\pi/2$, and therefore can be separated using a polarizing beam splitting (PBS) cube, thus realizing a narrow band atomic filter for Raman light.

\begin{figure}[lb]
\centerline{\includegraphics[]{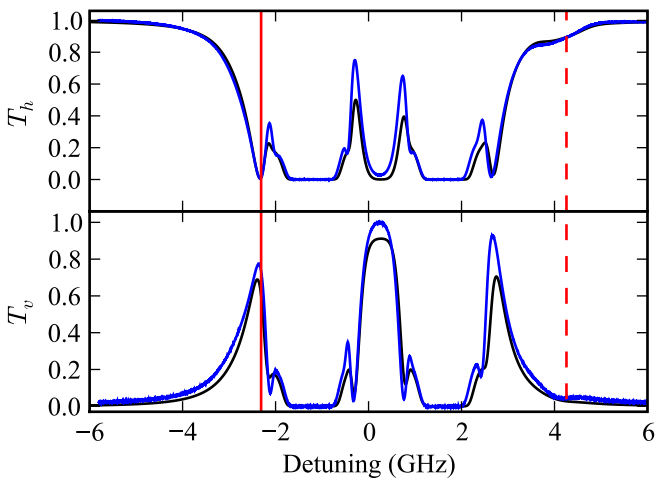}}
\caption{Transmission spectra at both output ports of the Faraday filter. The upper panel corresponds to the transmitted port of the filter with the bottom panel corresponding to the reflected port. The temperature of the filter is approximately 70$^{\circ}$~C and the magnetic field is 80~G. The blue line represents the experimental data and the black line is the result of a theoretical calculation. The solid red vertical line shows the frequency of the carrier and the dashed red line the frequency of the sideband, $\nu+\nu_m$, generated by the EOM. Zero detuning corresponds to the weighted center of the D${_2}$ line.   }
\label{fig:spectra}
\end{figure}

A schematic experimental setup of the Faraday filter is shown in Figure \ref{fig:setup}~(a). Light from a diode laser (Toptica DLPRO) is frequency stabilized to the $^{87}$Rb $5S_{1/2} (F=2) \rightarrow 5P_{3/2} (F'=3)$ transition using modulation transfer spectroscopy \cite{Mcca08} before passing through an EOM (New Focus 4851 driven by a Aglient E8267D signal generator) modulated at $\nu_m = 6.8$ GHz (corresponding to the ground state hyperfine-splitting in $^{87}$Rb as shown in Figure~\ref{fig:setup}~(b)). The EOM generates sidebands at $\pm\nu_m$ with an amplitude of 1\%. The beam is then linearly polarized by a PBS cube and propagated through a heated isotoptically pure $^{85}$Rb vapor cell (Triad Technology, TT-Rb85-75-V-Q) within a solenoid 28 cm in length with a field uniformity better than 1\% over the length of cell. The Rb vapor cell contains a mixture of Rb isotopes in a ratio of $^{85}$Rb : $^{87}$Rb equal to 99.9:0.1. The current in the solenoid produces both a homogenous magnetic field along the length of the cell and Ohmic heating, with additional temperature control provided by a Peltier effect heater. Following the vapor cell the light is analysed using a second PBS cube, which generates the two output ports of the filter.               

\begin{figure}[lb]
\centerline{\includegraphics[]{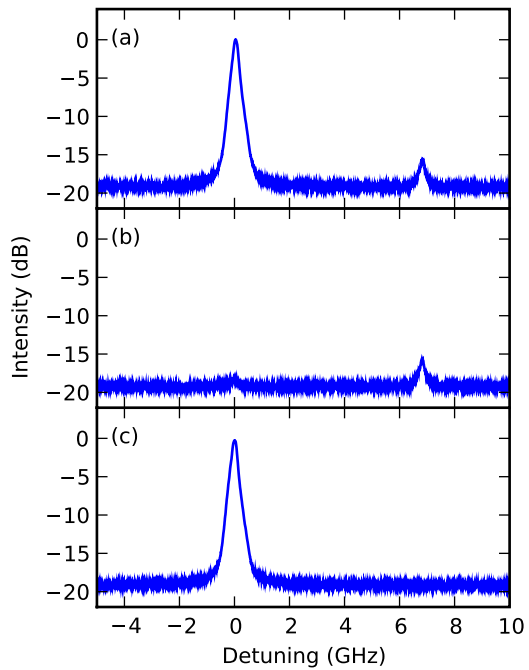}}
\caption{ Fabry-Perot etalon signals (a) with the filter turned off, (b) at the transmitted port of the polarizer ($T_h$) and (c) at the reflected port ($T_h$). In (a) both carrier and sideband are present, in (b) the carrier is suppressed to approximately -18 dB and in (c) the sideband is suppressed to the noise level leaving only the carrier.   }
\label{fig:filter_performance}
\end{figure}

The filter output as a function of the laser frequency for the Rb D$_2$ line is shown in Figure \ref{fig:spectra}. The detuning axis is scaled so that zero corresponds to the weighted center of the D${_2}$ line. The upper panel shows the transmission of light with a horizontal polarization, $T_h$, and the lower panel light with a vertical polarization, $T_v$. The experimental data are represented by the blue line and the black line is a theoretical calculation, which is seen to be in good agreement. The calculation is based on the theoretical model presented in \cite{Siddons08} and the polarization rotation found using equation \ref{equation:pol_rot} from \cite{Siddons09}. The magnetic field dependence is calculated using a full matrix diagonalization as in \cite{Popescu04}. The solid red vertical line indicates the position of light at the carrier frequency and the dashed red line shows the sideband frequency, $\nu+\nu_m$. The temperature of the filter is approximately 70$^{\circ}$C and the magnetic field generated by the solenoid is 80~G. The measurement is taken in the `weak probe' regime \cite{Sherlock09}. At the carrier frequency, $T_v$ is approximately 70\% whereas $T_h$ is close to zero. At the sideband frequency, $\nu+\nu_m$, the contrary is the case, with $T_h$ being approximately 90\% and $T_v$ is less than 1\%. This produces the desired separation of light at frequencies 6.8 GHz apart without a large portion of the light being absorbed by the vapor cell. The Verdet constant for the filter can be found using equation \ref{eq:verdet}. With a magnetic field of 80~G the Verdet constant $V=2600$ rad T$^{-1}$ m$^{-1}$, which is more than an order of magnitude larger than typical Faraday isolators.

In Figure~\ref{fig:filter_performance} the performance of the filter is demonstrated by using a Fabry-Perot etalon with a free spectral range $\sim 745$ MHz to analyse the light (a) with the magnetic field switched off, (b) at the transmitted port of the polariser ($T_h$) and (c) at the reflected port ($T_v$). The detuning axis is calibrated by taking the separation of the sidebands and carrier as being equal to the EOM modulation frequency. The second sideband, appearing at a detuning of -6.8~GHz, is omitted for clarity as it has no relevance to the Raman resonance. In Figure \ref{fig:filter_performance} (a) the two components of the input beam are clearly visible with the signal dominated by the carrier. In Figure \ref{fig:filter_performance} (b) the carrier frequency is suppressed to approximately -18 dB, leaving the sideband unchanged. In Figure \ref{fig:filter_performance} (c) the sidebands are suppressed to a level of -19 dB  leaving only the carrier. The suppression of the carrier is limited by the extinction ratio of the polarization optics used.

In summary, we have demonstrated the filtering of light separated in frequency by the ground state hyperfine splitting in $^{87}$Rb using the Faraday effect in an isotopically pure $^{85}$Rb thermal vapor. This technique allows the generation of phase stable beams suitable for driving Raman transitions between the two hyperfine states or the filtering of light generated by a Raman scattering process. The filter has been shown to result in a suppression of the carrier frequency to approximately -18 dB. We expect this work to be applicable to other alkali metal atom experiments where narrow band atomic filtering is required. 

The authors are grateful to S. L. Cornish for the loan of equipment. We thank the EPSRC for financial support.


\end{document}